\documentclass[a4paper, 12 pt, twoside, reqno]{amsart}

\usepackage[english]{babel}
\usepackage[utf8]{inputenc}

\usepackage{setspace}
\usepackage[euler-digits,euler-hat-accent]{eulervm}
\usepackage{times}
\usepackage{amsmath}
\usepackage{amsfonts}
\usepackage{amssymb}
\usepackage{amsmath}
\usepackage{amsthm}
\usepackage{graphicx}
\usepackage{array}
\usepackage{color}
\usepackage{mathrsfs}
\usepackage{hyperref}
\usepackage{eucal}
\usepackage{esint} %%% for \fint
\usepackage{tikz}
\usepackage{upgreek}
\usepackage{enumitem}

\usepackage{tikz-cd}

\allowdisplaybreaks

\theoremstyle{plain}
\newtheorem{theorem}{Theorem}[section]
\newtheorem{corollary}[theorem]{Corollary}
\newtheorem{proposition}[theorem]{Proposition}
\newtheorem{lemma}[theorem]{Lemma}

\theoremstyle{definition}

\theoremstyle{remark}
\newtheorem{remark}[theorem]{Remark}

\numberwithin{equation}{section}
\numberwithin{figure}{section}
\numberwithin{table}{section}

\newcommand{\N}{\mathbb{N}}
\newcommand{\C}{\mathbb{C}}                           

\newcommand{\Z}{\mathbb{Z}}

\newcommand{\s}[1]{\CMcal{#1}}
                  
\newcommand{\bb}[1]{\mathscr{#1}}
\newcommand{\rr}[1]{\mathfrak{#1}}
\newcommand{\n}[1]{\mathbb{#1}}

\newcommand{\dd}{\,\mathrm{d}}

\newcommand{\virg}[1]{\lq\lq#1\rq\rq}                \newcommand{\ie}{\textsl{i.\,e.\,}}

\DeclareMathOperator{\Ran}{Ran} 
\DeclareMathOperator{\Id}{Id}

\newlength{\dhatheight}

\hypersetup{
pdftoolbar=true,        % show Acrobat’s toolbar?
pdfmenubar=true,        % show Acrobat’s menu?
pdffitwindow=true,     % window fit to page when opened
pdfstartview=true,    % fits the width of the page to the window
pdftitle={Topology-of-States},    % title
pdfauthor={Giuseppe De Nittis, Santiago G. Rendel},     % author
breaklinks=true, %permet le retour a la ligne dans les liens trop longs
colorlinks=true,       % false: boxed links; true: colored links
linkcolor=purple,         % color of internal links (change box color 
citecolor=teal, % color of links to bibliography
urlcolor=blue, %couleur des hyperliens
bookmarksopen=true, %si les signets Acrobat sont crees,
filecolor=magenta,      % color of file links
}

\begin{document}

\title[Hilbert Grassmannians as classifying spaces]{
Hilbert Grassmannians as classifying spaces}

\author[G. De~Nittis]{Giuseppe De Nittis}

\address[G. De~Nittis]{Facultad de Matemáticas \& Instituto de Física,
  Pontificia Universidad Católica de Chile,
  Santiago, Chile.}
\email{gidenittis@uc.cl}

\author[K. Gomi]{Kiyonori Gomi}

\address[K. Gomi]{Department of Mathematics, Institute of Science Tokyo,
2-12-1 Ookayama, Meguro-ku, Tokyo, 152-8551, Japan.}
\email{kgomi@math.titech.ac.jp}

\author[S.~G. Rendel]{Santiago G. Rendel*}

% \address[S.~G. Rendel]{Facultad de Matemáticas \& Instituto de Física,
%   Pontificia Universidad Católica de Chile,
%   Santiago, Chile.}
% \email{sgr@uc.cl}
\address[S.~G. Rendel]{Department of Mathematics, KU Leuven, Celestijnenlaan 200b, box 2400, 3001 Leuven, Belgium.}
\email{santiago.g.rendel@kuleuven.be}

\vspace{2mm}

\date{\today\\
*Corresponding author}

\begin{abstract}
In this short work we prove that all standard operator topologies coincide on finite-rank Hilbert Grassmannians, and thus they are models for the classifying spaces of the finite-dimensional unitary groups. %As an application of this result, one can use Hilbert Grassmannians for the presentation of the $K$-theory of topological spaces by computing equivalence classes of homotopy equivalent maps valued in Hilbert Grassmannians.
As an application of this result, one can obtain the presentation of the $K$-theory of topological spaces by computing equivalence classes of homotopy equivalent maps valued in Hilbert Grassmannians.
\\
 \medskip

\noindent
{\bf MSC 2020}:
Primary: 	14M15;
Secondary: 	 55R35, 55R45, 46L30.\\
\noindent
{\bf Keywords}:
{\it Hilbert Grassmannians, weak topology, classifying space, $K$-theory.}
\end{abstract}

\maketitle

\tableofcontents

\section{Introduction}\label{sec:Intr0}

The \emph{classifying space} $B\n{G}$ of a topological group $\n{G}$ can be obtained as the quotient of a weakly contractible space $E\n{G}$ by a proper free action of $\n{G}$. Let us recall that the weak  contractibility of $E\n{G}$ means that all its homotopy groups are trivial, \ie $\pi_k(E\n{G})=0$ for every $k\in\N\cup\{0\}$. In particular $\pi_0(E\n{G})=0$ means that $E\n{G}$ is path-connected. The existence of the universal $\n{G}$-bundle $\n{G}\to E\n{G}\to B\n{G}$ is guaranteed by the \emph{Milnor construction} \cite{Milnor-56}.
The classifying space $B\n{G}$ is unique up to \emph{weak homotopy equivalence}, meaning that two classifying spaces for the same group $\n{G}$ are related by a continuous map which induces isomorphisms of all the homotopy groups. Additionally, if one insists that $B\n{G}$ must have a CW-complex structure (always possible up to a CW-approximation), then the classifying space is unique up to homotopy equivalence as a consequence of the \emph{Whitehead theorem}. 
Let $[X,B\n{G}]$ be the set of homotopy classes of continuous maps between the topological spaces $X$ and  $B\n{G}$. 
The defining property of $B\n{G}$ is that the set $[X,B\n{G}]$ provides the classification of principal $\n{G}$-bundles over $X$ up to isomorphism.
In the relevant case that  $X$ is (homotopy equivalent to) a CW-complex, $[X,B\n{G}]$ doesn't change under weak homotopy equivalence of the classifying space $B\n{G}$ \cite[Theorem 2]{matumoto-minami-sugawara-84}.  For more details about the theory of the universal $\n{G}$-bundle and classifying spaces we refer to \cite[Sections 3.4 \& 3.5]{rudolph-schmidt-17}.

 \medskip
  The classifying spaces of the unitary groups $\n{U}(n)$, with $n\in\N$  have a special role in the applications. In fact 
one has bijections of sets 
\begin{equation}\label{eq_class_vect_classic}
[X,B\n{U}(n)]\;\simeq\;{\rm Vec}_\C^n(X)\;\simeq\;{\rm Hilb}_{\rr{h}_n}(X)
\end{equation}
showing that $B\n{U}(n)$ provides the homotopy description of the set ${\rm Vec}_\C^n(X)$ of isomorphism classes of rank-$n$ complex vector bundles over $X$, or equivalently of ${\rm Hilb}_{\rr{h}_n}(X)$, the set of isomorphism classes of Hilbert bundles over $X$ with typical fiber the $n$-dimensional Hilbert space $\rr{h}_n$ \cite{dupre-74,schottenloher-18}. The classifying space $B\n{U}(\infty)$ of the \emph{infinite} unitary group $\n{U}(\infty)$ (the inductive limit of growing finite unitary groups, see Appendix \ref{app_unit_op}) provides the homotopy description of the set ${\rm Vec}_\C(X)$ of \emph{stable} equivalence classes of complex vector bundles over $X$, or equivalently of the reduced topological $K$-theory of $X$  \cite[Section 9.4]{husemoller-94}. In fact one has the bijection of sets
\begin{equation}\label{eq_class_vect_classic-02}
[X,B\n{U}(\infty)]\;\simeq\;{\rm Vec}_\C(X)\;\simeq\;\widetilde{K}^0(X)\;.
\end{equation}

 \medskip
 
 Although the existence of the spaces $B\n{U}(n)$ and $B\n{U}(\infty)$ is guaranteed by abstract arguments, in applications one is interested in concrete and manageable models representing these spaces.
 Requiring a CW- complex structure for $B\n{U}(n)$ and $B\n{U}(\infty)$ (always possible up to a CW-approximation), then, one has the \emph{identifications}
 $B\n{U}(n)\thickapprox {\rm Gr}_{n,\infty}$ and $B\n{U}(\infty)\thickapprox{\rm Gr}_{\infty}$ where  and 
 ${\rm Gr}_{n,\infty}$ and ${\rm Gr}_{\infty}$ are inductive limits of the  classical Grassmannians ${\rm Gr}_{n}(\C^d)$ (see Appendix \ref{sec:class_Grass} for more details). By identification $\thickapprox$ we mean a \emph{homotopy equivalence} and we will denote with the symbol $\thicksim$ a \emph{weak homotopy equivalence}. 
By insisting on the CW-complex structure of the classifying spaces, we will freely use the classical spaces ${\rm Gr}_{n,\infty}$ and ${\rm Gr}_{\infty}$ as models for $B\n{U}(n)$ and $B\n{U}(\infty)$, respectively.

\medskip

In the study of topological phenomena emerging in quantum theories 
one is led to construct models of the  classifying spaces built inside the framework of a separable infinite dimensional Hilbert space. Henceforth, we will denote with $\rr{h}$ 
 a separable complex Hilbert space of dimension ${\rm dim}(\rr{h})=\aleph_0$ and with $\bb{B}(\rr{h})$ its von Neumann algebra of bounded operators. 
 One interesting example of classifying space defined inside the Hilbert space theory is provided by the 
 \emph{Atiyah-J\"{a}nich Theorem} \cite[Theorem A1]{Atiyah-67}  which states that the set $\bb{F}(\rr{h})\subset\bb{B}(\rr{h})$ of (bounded) Fredholm
operators on $\rr{h}$, endowed  with the uniform topology of the operator norm, is a classifying space for the $K$-theory. More precisely one gets that $[X,\bb{F}(\rr{h})]\simeq K^0(X)$ 
 where $X$ is any compact space. This shows in particular that $\bb{F}(\rr{h})\sim B\n{U}(\infty)\times \Z$ (where $\Z$ is regarded as a discrete topological space) \cite[Section 9.4, Remark 4.6]{husemoller-94}. Another example of  classifying space for $K$-theory
  builded inside the Hilbert space framework is discussed \cite[Section 5]{kuiper-65}.

 \medskip
 
 Our interest consists in the use of \emph{Hilbert Grassmannians} for the construction of models of classifying spaces.
 Let us introduce the group of the unitary operators by $\bb{U}(\rr{h})\subset \bb{B}(\rr{h})$, and the complete lattice of (orthogonal) projections 
 $\bb{P}(\rr{h})\subset \bb{B}(\rr{h})$, \ie of the elements $P\in\bb{B}(\rr{h})$ such that $P^*=P=P^2$. Let ${\rm Tr}_{\rr{h}}$ be the canonical trace on $\rr{h}$. A projection $P\in\bb{P}(\rr{h})$ has finite rank $n\in\N$ when ${\rm Tr}_{\rr{h}}(P)=n$.
The Banach space of  trace-class operators 
will be denoted with $\bb{L}^1(\rr{h})$.
Following \cite{shubin-96,abbondandolo-majer-09}, 
one can introduce the Hilbert Grassmannian of rank  $n\in\N$ as the set%
\begin{equation}\label{eq:gra_1k}
\bb{G}_n(\rr{h})\;:=\;\left\{P\in\bb{P}(\rr{h})\;|\; {\rm Tr}_{\rr{h}}(P)=n \right\}\;,
\end{equation}
\ie as the space of orthogonal projections of $\rr{h}$ with rank $n$.
It  will be useful to denote with $\bb{G}_0(\rr{h}):=\{0\}$ the singleton consisting of the \emph{null-projection}.
Any element $P\in \bb{G}_n(\rr{h})$ can be  uniquely identified with a closed subspace $\rr{v}_P:={\rm Ran}(P)\subset\rr{h}$ of dimension $n$ and this identification is bijective. As a consequence one can interpret $\bb{G}_n(\rr{h})$ as the set of the $n$-dimensional closed subspaces of $\rr{h}$. 
We also need the set
\begin{equation}\label{eq:gra_1k-33}
\bb{G}_\bullet(\rr{h})\;:=\;\bigcup_{n=0}^\infty\bb{G}_n(\rr{h})\; =\;\bb{P}(\rr{h})\cap \bb{L}^1(\rr{h})
\end{equation}
consisting of all trace-class projections. The spaces $\bb{G}_n(\rr{h})$ and $\bb{G}_\bullet(\rr{h})$ can be endowed with all the relevant topologies that one can define on $ \bb{B}(\rr{h})$. One natural choice is to use the \emph{uniform topology} of the operator norm  
 and in this case we will denote the related spaces with ${^{\rm u}\bb{G}_n(\rr{h})}$ and ${^{\rm u}\bb{G}_\bullet(\rr{h})}$.
 This is the situation studied in \cite{abbondandolo-majer-09}.
A different, less restrictive option is to endow the Hilbert Grassmannians with the \emph{weak topology} of operators. The resulting spaces will be denoted with  ${^{\rm w}\bb{G}_n(\rr{h})}$ and ${^{\rm w}\bb{G}_\bullet(\rr{h})}$ in this work. 
The weak topology of the Hilbert Grassmannians has been extensively studied in \cite{shubin-96}. A relevant fact that is worth mentioning right away is that on these spaces the strong operator topology agrees with the weak operator topology \cite[Proposition 3.1]{shubin-96}. Our main Theorem below asserts that actually most operator topologies one usually considers coincide in these spaces. More background information about the topology of Hilbert Grassmannians will be provided in Section \ref{hil_gras}.

\medskip

In the following let us use the notation $\N_0:=\N\cup\{0\}$. The main achievement  of this work is contained in the following result, which will be covered in Section \ref{sec:fin-R}.
\begin{theorem}[Hilbertian classifying spaces]\label{thm:main}
The (ultra)weak, (ultra)strong, norm, and Schatten $p$-norm with $p\geq 1$ topologies all coincide on $\bb{G}_n(\rr{h})$. 
In particular, the following homotopy equivalences hold true:
\begin{equation}%\label{eq:main_01}
	B\n{U}(n)\;\thickapprox\;{^{\rm u}\bb{G}_n(\rr{h})}\;\thickapprox\;{^{\rm w}\bb{G}_n(\rr{h})}\;,\quad \forall\; n\in\N_0\;.
\end{equation}
\end{theorem}

\medskip

The first homotopy equivalence is known by \cite[Section 2.3]{abbondandolo-majer-09}, together with the fact that $GL(n)\approx \mathbb{U}(n)$ (indeed, $\mathbb{U}(n)$ is a deformation retract of $GL(n)$), and thus $BGL(n)\approx B\mathbb{U}(n)$. The remaining equivalences are new, at least to the best of our knowledge. 
To obtain the first equivalence, the authors employ a relevant result on the structure of ${^{\rm u}\bb{G}_n(\rr{h})}$, recovering it as a quotient of the unitary group ${^{\rm u}\bb{U}(\rr{h})}$. More precisely, fix a projection $P_0 \in \bb{G}_n(\rr{h})$, and denote its range by $\rr{v}_0$. Then the map $\pi:{^{\rm u}\bb{U}(\rr{h})} \to {^{\rm u}\bb{G}_n(\rr{h})}$ defined by $\pi:U\mapsto UP_0U^*$ is a locally trivial fiber bundle, with typical fiber ${^{\rm u}\bb{U}(\rr{v}_0)}\times {^{\rm u}\bb{U}(\rr{v}_0^\bot)}$, which corresponds to the isotropy group of $P_0$ under the action of ${^{\rm u}\bb{U}(\rr{h})}$ by conjugation. In particular, this provides a homeomorphism
\[
{^{\rm u}\bb{G}_n(\rr{h})}\;\simeq\;{^{\rm u}\bb{U}(\rr{h})}/\left[{^{\rm u}\bb{U}(\rr{v}_0)}\times {^{\rm u}\bb{U}(\rr{v}_0^\bot)}\right]\;.
\]
This can then be used, for instance, to prove a weak homotopy equivalence ${^{\rm u}\bb{G}_n(\rr{h})}\sim B\mathbb{U}(n)$, which then lifts to the desired homotopy equivalence thanks to \cite[Theorem 15]{palais-66}. 
As a secondary result, we abstain from using the homeomorphisms presented in Theorem \ref{thm:main} and, focusing solely on the weak topology, we recover this structural result for the weak Grassmannians and weak unitary groups:
\begin{proposition}\label{prop:wheq}
Let $n\in\N$, $P_0\in \bb{G}_n(\rr{h})$, and denote by $\rr{v}_0$ the range of ${P_0}$. The map %$\pi:{^{\rm w}\bb{U}(\rr{h})} \to {^{\rm w}\bb{G}_n(\rr{h})}$ defined by $\pi:U\mapsto UP_0U^*$ 
\begin{align*}
    \pi:{^{\rm w}\bb{U}(\rr{h})} &\to {^{\rm w}\bb{G}_n(\rr{h})}\\
    U&\mapsto UP_0U^*
\end{align*}
is a locally trivial fiber bundle, with typical fiber ${^{\rm w}\bb{U}(\rr{v}_0)}\times {^{\rm w}\bb{U}(\rr{v}_0^\bot)}$. In particular,
\begin{equation}\label{eq:gra_set-w-fin}
{^{\rm w}\bb{G}_n(\rr{h})}\;\simeq\;{^{\rm w}\bb{U}(\rr{h})}/\left[{^{\rm w}\bb{U}(\rr{v}_0)}\times {^{\rm w}\bb{U}(\rr{v}_0^\bot)}\right]\;.
\end{equation}
\end{proposition}
This result allows us to reconstruct ${^{\rm w}\bb{G}_n(\rr{h})}$ as a quotient of the weak unitary group of $\rr{h}$, and to directly obtain the weak homotopy equivalence ${^{\rm w}\bb{G}_n(\rr{h})}\sim B\mathbb{U}(n)$.

\medskip

The search for a model of the classifying space $B\n{U}(\infty)$ in terms of Hilbert Grassmannians is less immediate. The first immediate candidates would seem to be the spaces ${^{\rm u}\bb{G}_\bullet(\rr{h})}$ and ${^{\rm w}\bb{G}_\bullet(\rr{h})}$. However, none of these spaces can be homotopy equivalent to $B\n{U}(\infty)$. In the first case the uniform topology turns out to be \virg{too strong} (Lemma \ref{lemm_nogo}) and in the second case the weak topology turns out to be \virg{too weak} (Corollary \ref{cor_nogo}).

\medskip

A relevant consequence of Theorem \ref{thm:main} is the possibility 
of using Hilbert Grassmannians with the weak topology as classifying spaces.

\begin{corollary}\label{co_VetB}
Let $X$ be a paracompact Hausdorff space
with the  homotopy type of a CW-complex. Then
\[
[X,{^{\rm w}\bb{G}_n(\rr{h})}]\;\simeq\;{\rm Vec}_\C^n(X)\;,
\]
for every $n\in\N_0$.
\end{corollary}

\medskip

The proof of this result is obtained by combining the homotopy equivalences in Theorem \ref{thm:main} with equations \eqref{eq_class_vect_classic} and \eqref{eq_class_vect_classic-02}
and the support of \cite[Theorem 2]{matumoto-minami-sugawara-84}.

\begin{remark}[Chern classes]
By combining the case $n=1$ with \eqref{eq_class_vect_classic2} one gets
 \[
[X,{^{\rm w}\bb{G}_1(\rr{h})}]\;\stackrel{c_1}{\simeq}\;H^2(X,\Z)\;.
\]
where the map $c_1$ is the \emph{first Chern class}.
\hfill $\blacktriangleleft$
\end{remark} 
 
 \begin{remark}[Reduced $K$-theory]
Let $X$ be a paracompact Hausdorff space
with the  homotopy type of a finite $CW$-complex of dimension $d\leqslant 3$. Then, one has that 
\[
[X,{^{\rm w}\bb{G}_n(\rr{h})}]\;\simeq\;\widetilde{K}^0(X)
\]
where $\widetilde{K}^0(X)$  denotes the reduced topological complex $K$-group of $X$.
This fact relies on  the \emph{stable range theorem} \cite[Theorem 1.2, Chapter 9]{husemoller-94} which implies that   the classification of vector bundles of any rank reduces to the classification of line bundles if the dimension of $X$ is bounded by $3$. This means that ${\rm Vec}_\C^n(X)\simeq {\rm Vec}_\C^1(X)$ for every rank $n$, and under the canonical inclusion ${\rm Vec}_\C^n(X)\hookrightarrow {\rm Vec}_\C^{n+1}(X)$ given by the sum   
of a trivial line bundle one gets ${\rm Vec}_\C^\bullet(X) \simeq {\rm Vec}_\C^1(X)$ where ${\rm Vec}_\C^\bullet(X)$ is the set of equivalence classes of complex vector bundles under \emph{stable} isomorphisms. 
However, the latter is classified exactly by the reduced $K$-theory of $X$, \ie ${\rm Vec}_\C^\bullet(X)\simeq \widetilde{K}^0(X)$.
\hfill $\blacktriangleleft$
\end{remark}

 \medskip
 
The physical relevance of the use of the  Hilbert Grassmannians with the {weak topology} as classifying spaces relies in the following construction. Let $\bb{A}\subseteq\bb{B}(\rr{h})$ a unital $C^*$-algebra
 and $P:X\to {^{\rm w}\bb{G}_\bullet(\rr{h})}$ a continuous function such that $1\leqslant{\rm Tr}_{\rr{h}}(P(x))\leqslant M$ for every $x\in X$ and some $M\in\N$  and such that the map $x\mapsto {\rm Tr}_{\rr{h}}(P(x))$ is continuous. Consider the family of states of $\bb{A}$ parametrized by $x\in X$ and defined by
 \[
 \omega_x(A)\;:=\;\frac{{\rm Tr}_{\rr{h}}(P(x)A)}{{\rm Tr}_{\rr{h}}(P(x))}\;,\qquad A\in \bb{A}\;.
 \]
 \begin{proposition}\label{prop:con_stat}
 The map $x\mapsto \omega_x$ is continuous with respect to the weak-$\ast$ topology of states.
 \end{proposition}
 
 \medskip
 
 This result suggests that the weak topology on the Hilbert Grassmannian is sufficient to guarantee continuity of the associated states with respect to the point-wise evaluation on observables. The proof is postponed to Section \ref{sec:pat_sta}.

\medskip

 \noindent
{\bf Structure of the paper.}
In {\bf Section~\ref{hil_gras}}, we present basic results and the general structure of Hilbert Grassmannians. Then, {\bf Section~\ref{sec:fin-R}} deals with the topology of Grassmannians of fixed dimension and the ingredients for the proofs of Theorem \ref{thm:main} and Proposition \ref{prop:wheq}. In {\bf Section~\ref{sec:infin-R}} we deal with the space of all trace-class projections, and prove no-go results showing that neither the norm nor weak topology endow it with the homotopy type of $B\n{U}(\infty)$. In {\bf Section~\ref{sec:pat_sta}} we prove Proposition \ref{prop:con_stat}. 
{\bf Appendix~\ref{sec:class_Grass}} provides a succinct exposition on the classical Grassmannians. In {\bf Appendix~\ref{app_unit_op}} we present the infinite unitary groups and their homotopy groups. {\bf Appendix~\ref{unif_gras}} shows the homotopy equivalence between the classical Grassmannians and the uniform Hilbert Grassmannians. Finally, in {\bf Appendix~\ref{weak_contr}} we construct the tools needed for the proof of Proposition \ref{prop_nogo-2}.

 \medskip
 
 \noindent
{\bf Acknowledgements.}
GD's research is supported by the grant \emph{Fondecyt Regular - 1230032}. KG's research is supported by \emph{JSPS KAKENHI} Grant Numbers 20K03606. SR would like to thank Manish Kumar for the stimulating discussions during the revision of this paper.

\section{Generalities on Hilbert Grassmannians}\label{hil_gras}
This section contains results from \cite{shubin-96,abbondandolo-majer-09}. Let $\rr{h}$ be a separable and infinite dimensional Hilbert space, \ie  ${\rm dim}(\rr{h})=\aleph_0$.
For every $n\in\N$ the space $\bb{G}_n(\rr{h})$ is defined by \eqref{eq:gra_1k} with the convention $\bb{G}_0(\rr{h})=\{0\}$.
The \emph{reciprocal} {Grassmannian} is
\begin{equation}\label{eq:gra_1k_recip}
\bb{G}_n^\bot(\rr{h})\;:=\;\left\{P^\bot:=({\bf 1}-P)\;|\; P\in \bb{G}_n(\rr{h})  \right\}\;
\end{equation}
with the convention $\bb{G}_0^\bot(\rr{h})=\rr{h}$.
Finally, the \emph{(purely) infinite} {Grassmannian} is 
\begin{equation}\label{eq:gra_1k_infinf}
\bb{G}_{\infty,\infty}(\rr{h})\;:=\;\left\{P\in\bb{P}(\rr{h})\;|\; P,P^\bot\notin \bb{L}^1(\rr{h})  \right\}\;.
\end{equation}
meaning that both $P$ and $P^\bot$ project on infinite-dimensional subspaces.
With the definition of $\bb{G}_\bullet(\rr{h})$ given by \eqref{eq:gra_1k-33}, and its dual
\[
 \bb{G}_\bullet^\bot(\rr{h})\;:=\;\bigcup_{k=0}^\infty\bb{G}_k^\bot(\rr{h})\;,
\]
one obtains the following partition of the set of projections of $\rr{h}$:
\[
\bb{P}(\rr{h})\;=\;\bb{G}_\bullet(\rr{h})\;\sqcup\;\bb{G}_\bullet^\bot(\rr{h})\;\sqcup\;\bb{G}_{\infty,\infty}(\rr{h})\;.
\]
It is worth commenting, as recalled in \cite[eq. (1.1)]{shubin-96}, that there is a bijection of sets
\begin{equation}\label{eq:gra_set}
\bb{G}_n(\rr{h})\;\simeq\;\bb{U}(\rr{h})/\left[\bb{U}(\rr{v}_0)\times \bb{U}(\rr{v}_0^\bot)\right]
\end{equation}
where the $n$-dimensional subspace $\rr{v}_0\subset\rr{h}$ is the range of any reference projection $P_0\in \bb{G}_n(\rr{h})$ and 
$\rr{v}_0^\bot$ is its orthogonal complement.  The set $\bb{U}(\rr{h})\subset \bb{B}(\rr{h})$ denotes the group of unitary operators on $\rr{h}$ (see Appendix \ref{app_unit_op} for more details), and $\bb{U}(\rr{v}_0)$ and $\bb{U}(\rr{v}_0^\bot)$ are the \emph{restriction} to 
$\rr{v}_0$ and $\rr{v}_0^\bot$, respectively. A similar bijection holds true for $\bb{G}_{\infty,\infty}(\rr{h})$.

\medskip

All the spaces above are subspaces of $\bb{B}(\rr{h})$ and in turn can be 
topologized with any of the topologies defined on  $\bb{B}(\rr{h})$. There are at least three relevant topologies: the norm (or uniform) topology, the strong topology and the weak topology. However, the weak and the strong topology agree on any of the 
Grassmannians above and on the full space $\bb{P}(\rr{h})$ \cite[Proposition 3.1]{shubin-96} (as well as on $\bb{U}(\rr{h})$ as reported in  Appendix \ref{app_unit_op}). Let us also observe that the mapping $P\mapsto {\bf 1}-P$ induces a homeomorphism between  $\bb{G}_n(\rr{h})$ and $\bb{G}^\bot_n(\rr{h})$ in any of the aforementioned topologies and for every $n\in\N$.
For this reason we will consider only the Grassmannians $\bb{G}_n(\rr{h})$ and $\bb{G}_{\infty,\infty}(\rr{h})$. We will use the symbols ${^{\rm u}}\bb{G}_n(\rr{h})$ and ${^{\rm w}}\bb{G}_n(\rr{h})$ for the Grassmannian endowed with the norm topology and the weak (or strong) topology, respectively. Likewise for the purely infinite Grassmannian.

\medskip

With respect to the norm topology one has that ${^{\rm u}}\bb{G}_n(\rr{h})$, ${^{\rm u}}\bb{G}_n^\bot(\rr{h})$ with $n\in\N$ and ${^{\rm u}}\bb{G}_{\infty,\infty}(\rr{h})$ are closed connected components of ${^{\rm u}}\bb{P}(\rr{h})$ \cite[Lemma 1.1 \& Corollary 1.3]{shubin-96}. Moreover \cite[Corollary 1.4]{shubin-96} implies that
\begin{equation}\label{eq:gra_set-u}
{^{\rm u}\bb{G}_n(\rr{h})}\;\simeq\;{^{\rm u}\bb{U}(\rr{h})}/\left[{^{\rm u}\bb{U}(\rr{v}_0)}\times {^{\rm u}\bb{U}(\rr{v}_0^\bot)}\right]
\end{equation}
as topological spaces, and the same for  ${^{\rm u}\bb{G}_{\infty,\infty}(\rr{h})}$.
The homotopy of these spaces is computed in \cite[Section 2.3]{abbondandolo-majer-09} as a direct consequence of equation \eqref{eq:gra_set-u} and Kuiper’s theorem that shows the contractibility of ${^{\rm u}\bb{U}(\rr{h})}$ when $\rr{h}$ is infinite dimensional.
 One has that ${^{\rm u}}\bb{G}_{\infty,\infty}(\rr{h})$ is a weakly contractible space, and in turn contractible in view of \cite[Theorem 15]{palais-66}. Moreover the {Grassmannians} ${^{\rm u}}\bb{G}_n(\rr{h})$ have the same homotopy type of the classifying space of $\n{U}(n)$, namely
 $B\n{U}(n)\sim{^{\rm u} \bb{G}_n(\rr{h})}$. To see this
 choose an orthonormal basis $\{e_j\}_{j\in\N}\subset\rr{h}$ and identify $\C^d$ with ${\rm span}(e_1,\ldots,e_d)\subset \rr{h}$.
This provides  inclusions $\jmath_d:{\rm Gr}_{n}(\C^d)\hookrightarrow {^{\rm u}\bb{G}_n(\rr{h})}$ which commutes with the inclusions ${\rm Gr}_{n}(\C^d)\subset {\rm Gr}_{n}(\C^{d+1})$, and therefore extends to the inductive limit
\begin{equation}\label{eq:incl_col}
\jmath\;:\;{\rm Gr}_{n,\infty}\;\hookrightarrow\; {^{\rm u}\bb{G}_n(\rr{h})}\;.
\end{equation}
The map $\jmath$ in \eqref{eq:incl_col} results to be continuous by construction and provides a weak homotopy equivalence. Although this result is claimed in the literature (see \cite[p. 28]{abbondandolo-majer-09} for instance), we were unable to find an explicit reference to this result. For that we will add an explicit proof in
 Appendix \ref{unif_gras} for the benefit of the reader. 

\medskip

With respect to the weak topology one has that ${^{\rm w}}\bb{G}_k(\rr{h})$, ${^{\rm w}}\bb{G}_k^\bot(\rr{h})$ with $k\in\N$ and ${^{\rm w}} \bb{G}_{\infty,\infty}(\rr{h})$ are
metrizable and separable, and the same holds true also for ${^{\rm w}}\bb{P}(\rr{h})$ \cite[Corollaries 2.5 \& 2.7]{shubin-96}. However these spaces are not closed in $\bb{B}(\rr{h})$ with respect to the weak topology. A description of the closures is provided in  
\cite[Propositions 3.3, 3.4 \& 3.5]{shubin-96}.

%----%

\section{The fixed rank case}\label{sec:fin-R}

To prove Theorem \ref{thm:main}, we will first prove the following proposition, and obtain it as a corollary. We will employ a strategy previously used to prove \cite[Theorem 5.1]{denittis-rendel-25} and \cite[Lemma 2.3]{denittis-25}. Let $\bb{L}^p(\rr{h})\subset \bb{B}(\rr{h})$ denote the $p$-\emph{Schatten ideal}, and $\|\cdot\|_{\rm p}$ the \emph{Schatten $p$-norm}. For $p=1$, this is just the trace-norm. Let $\bb{L}^p_{+,1}(\rr{h})\subset \bb{L}^p(\rr{h})$ be the subset of positive elements $A\in \bb{L}^p(\rr{h})$ with $\|A\|_{\rm p} = 1$. In particular, $\bb{L}^1_{+,1}(\rr{h})$ denotes the positive trace-class operators with trace equal to one.
\begin{proposition}\label{prop:L1_top}
        The weak and Schatten $p$-norm topologies are sequentially homeomorphic on $\bb{L}^p_{+,1}(\rr{h})$, for all $1\leq p < \infty$. In particular, all the topologies mentioned in Theorem \ref{thm:main} are sequentially homeomorphic on $\bb{L}^1_{+,1}(\rr{h})$.
\end{proposition}
\proof  Let $\{A_i\}_{i\in \N}$ be a sequence in $\bb{L}^p_{+,1}(\rr{h})$ and $A$ an element. Any Schatten $p$-norm topology is always stronger than the weak topology: If $\|A_i-A\|_{\rm p}\to 0$, then $\|A_i-A\|\leq \|A_i-A\|_{\rm p}\to 0$, so $A_i\to A$ in norm, and thus weakly. Let us prove the converse in the specified space. Suppose $A_i \to A$ weakly. Since $A_i = A_i^* = |A_i|$ and same for $A$, and $\|A_i\|_{\rm p} = \|A\|_{\rm p} = 1$, we can use \cite[Theorem 2.20]{simon-05} to deduce that in fact $\|A_i-A\|_{\rm p}\to 0$.
\qed

\begin{proof}[Proof of Theorem \ref{thm:main}]
    By scaling by $n^{-1/p}$, we can embed $\bb{G}_n(\rr{h})$ into $\bb{L}^p_{+,1}(\rr{h})$, so by \ref{prop:L1_top}, the identity map ${^{\rm 1}\bb{G}_n(\rr{h})} \to {^{\rm w}\bb{G}_n(\rr{h})}$ is a sequential homeomorphism. On the other hand, by \cite[Corollary 2.5]{shubin-96}, ${^{\rm w}\bb{G}_n(\rr{h})}$ is metrizable. Since we have a sequential homeomorphism between two metrizable, hence sequential spaces, it turns out to be a genuine homeomorphism.
\end{proof}

\medskip

The above theorem, combined with the homotopy equivalence ${^{\rm u}\bb{G}_n(\rr{h})} \approx B\mathbb{U}(n)$, provides the desired homotopy equivalence ${^{\rm w}\bb{G}_n(\rr{h})} \approx B\mathbb{U}(n)$. As was mentioned in the introduction, the first of these equivalences is obtained through the use of the additional structure described in Equation \ref{eq:gra_set-u}, and the associated fibration. We will now prove Proposition \ref{prop:wheq}, which asserts that the same structure is present when we replace the uniform topology by the weak topology. 

\begin{proof}[Proof of Proposition \ref{prop:wheq}]
To prove this, one can adapt the argument in \cite[Proposition 1.2]{shubin-96}.
Step $1^o$ is checking the continuity of $\pi$.
First of all since both $^{\rm w}\bb{U}(\rr{h})$ and $^{\rm w}\bb{G}_k(\rr{h})$ are metrizable spaces the continuity of the map $\pi$ can be controlled on sequences.
Then the continuity of $\pi$ follows by the inequality
\[
|\langle f,(UP_0U^*-P_0) g\rangle|\;\leqslant\;|\langle P_0U^*f,(U^*-{\bf 1}) g\rangle|
+|\langle f,(U-{\bf 1}) P_0g\rangle|
\]
valid for every $f,g\in\rr{h}$
which shows that when $U\to{\bf 1}$ in the weak topology then also $\pi(U)=UP_0U^*\to P_0=\pi(\bf 1)$ weakly. The most general case follows by the same argument by observing that
\[
\langle f,(UP_0U^*-VP_0V^*) g\rangle\;=\;\langle Vf,(V^*UP_0U^*V-P_0) V^*g\rangle\]
for every pair of unitaries $U$ and $V$. What it remains to prove is the existence of local trivializations.
By repeating almost verbatim step $2^o$ of \cite[Proposition 1.2]{shubin-96} this is equivalent to the existence of local sections. In fact the only difference in the original argument is that continuity of the local trivialization defined by the local section follows from the fact that ${^{\rm w}\bb{U}}(\rr{h})$ is a topological group also in the weak topology \cite[Theorem 1.2]{espinoza-uribe-14}.
To conclude the proof according to step $3^o$ of \cite[Proposition 1.2]{shubin-96}, one has to show that for any $Q \in {^{\rm w}\bb{G}_k}(\rr{h})$ we have a neighbourhood $\s{O}$ of $Q$ and a continuous map $s:\s{O} \to \pi^{-1}(\s{O})$ such that $\pi\circ s = \Id_\s{O}$ (a local section).
First let $Q = P_0$  and fix an orthonormal basis of $\rr{v}_0$ given by $\{b_1,\ldots,b_k\}$. 
With this, one can define a neighbourhood of $P_0$ given by
\[
\begin{aligned}
\s{O}_{0}\;:&=\; \left\{ P\in {^{\rm w}\bb{G}_k(\rr{h})} \; \big| \; |\langle b_i, (P_0- P) b_j\rangle| < \varepsilon_k^2,\;\; i,j=1,\dots,k \right\}\\
&=\;\left\{ P\in {^{\rm w}}\bb{G}_k(\rr{h}) \; \big| \; \|(P_0  - P )b_j\| < \varepsilon_k,\; j=1,\dots,k \right\}\;.
\end{aligned}
\]
The second equality follows by observing that
\[
\langle b_i, (P_0 - P) b_j\rangle\; =\; \langle b_i, ({\bf 1} - P) b_j\rangle\; =\; \langle b_i, ({\bf 1} - P)^2 b_j\rangle\; =\; \langle b_i, (P_0 - P)^2 b_j\rangle
\]
implies $|\langle b_j, (P_0- P) b_j\rangle|=\|(P_0- P) b_j\|^2$ for every $j=1,\ldots,k$. This fact, along with the Cauchy-Schwarz inequality
\[
|\langle b_i, (P_0- P) b_j\rangle| \;=\; |\langle (P_0- P) b_i, (P_0- P) b_j\rangle| \;\leqslant\; \|(P_0- P) b_i\|\|(P_0- P) b_j\|
\]
for every $i,j=1,\ldots,k$, provides the equivalence of the two descriptions of $\s{O}_0$. 
The number $\varepsilon_k>0$,  will be assumed initially sufficiently small (depending on $k$) and we will fix a sufficient explicit bound during the proof.
In order to define a local section on  $\s{O}_0$ let us introduce the
operator $A_P:= P P_0 + P^\perp P_0^\perp$, where $P^\perp:={\bf 1}-P$
and $P^\perp_0:={\bf 1}-P_0$. Given its polar decomposition $A_P = U_P|A_P|$,
we will define our section on $\s{O}_0$ as the map $s_0:P\mapsto U_P$.
The crucial point is to prove that $U_P$ is indeed a unitary operator, continuous on $P$, and such that $U_P P_0 U_P^* = P$. As in step $3^o$ of \cite[Proposition 1.2]{shubin-96} this follows by proving that 
$A_P$ is invertible and defines linear topological isomorphisms $\rr{v}_0\to\rr{v}$ and $\rr{v}_0^\perp\to\rr{v}^\perp$
where $\rr{v}:=\Ran(P)$. We will leave the proof of the continuity of $\s{O}_0 \ni P\mapsto U_P = A_P |A_P|^{-1} \in {^{\rm w}}\bb{B}(\rr{h})$ for last.
The remaining properties follow as in step
$3$° of \cite[Proposition 1.2]{shubin-96}.
Note that $P\in \s{O}_0$ implies
$\|P^\perp b_j\| < \varepsilon_k$ for every $j=1,\ldots,k$.
Let $f\in \rr{v}_0 \setminus \{0\}$. Then
\begin{equation*}\label{eq:proof_PperpP0}
    \| P^\perp f\|^2\; =\; \left\|\sum_{j=1}^k f_j P^\perp b_j \right\|^2 \;\leq\;  \|f\|^2 \sum_{j=1}^k \|P^\perp b_j\|^2\; <\;  k\varepsilon_k^2\|f\|^2\;,
\end{equation*}
or equivalently $\|P^\perp P_0\| < \sqrt{k}\varepsilon_k$. Let us fix $\sqrt{k}\varepsilon_k <1$.
Observing that 
\[
\|f\|^2\;=\;\|Pf+P^\perp f\|^2\;=\;\|Pf\|^2+\|P^\perp f\|^2\;<\;k\varepsilon_k^2\|f\|^2+\|P f\|^2
\]
one gets that $\|P f\|^2>c_k^2\|f\|^2$ for every $f\in \rr{v}_0 \setminus \{0\}$ with $c_k^2:=1-k\varepsilon_k^2>0$. This means that $PP_0$ is an injective map from $\rr{v}_0$ to $\rr{v}$, and in view of the fact that the two spaces have  same dimension $k$, it follows that $A_{P,0} :=PP_0|_{\rr{v}_0}:\rr{v}_0\to\rr{v}$ is a linear isomorphism (hence invertible) between $\rr{v}_0$ and $\rr{v}$, and by definition $A_P|_{\rr{v}_0} =A_{P,0}$. 
Note that $\|A_{P,0}^{-1}\|<c_k^{-1}$ in view of the inequality above. From
\begin{equation}\label{eq_intr_calc}
\langle g , A_{P,0} f \rangle_{\rr{v}} \;=\; \langle g , PP_0 f \rangle_{\rr{h}} \;=\; \langle P_0 P g , f \rangle_{\rr{h}} \;=\; \langle P_0 P g , f \rangle_{\rr{v}_0}
\end{equation}
valid for every $f\in \rr{v}_0$ and $g\in \rr{v}$ one infers that 
$A_{P,0}^*:=P_0 P|_{\rr{v}}:\rr{v} \to \rr{v}_0$. The invertibility of $A_{P,0}$ implies the invertibility of $A_{P,0}^*$, and by standard equalities one gets that 
$\|(A_{P,0}^*)^{-1}\| = \|(A_{P,0}^{-1})^*\| = \|A_{P,0}^{-1}\|<c_k^{-1}$.
This implies that 
\[
\begin{aligned}
\|P_0^\perp P g\|^2 \;&=\; \|P g\|^2 - \|P_0 P g\|^2 \;=\;  \|g\|^2 - \|A_{P,0}^* g\|^2\\ &\leqslant\; \|g\|^2 - \frac{\|g\|^2}{\|(A_{P,0}^*)^{-1}\|^2} \;<\; (1-c_k^2)\|g\|^2
\end{aligned}
\]
for every 
$g\in \rr{v}$. Let $C_P:=P_0^\perp P|_{\rr{v}}:\rr{v}\to\rr{v}_0^\perp$
and from the same argument in \eqref{eq_intr_calc} observe that $C_P^*=PP_0^\perp|_{\rr{v}_0^\perp}$. From $\|C_P\|^2=\|C_P^*\|^2<(1-c_k^2)$ one infers
\begin{equation}\label{eq:proof_Rpz_lowerbound}
\|P^\perp P_0^{\perp} h\|^2 \;=\; \|P_0^{\perp} h\|^2 - \|P P_0^{\perp} h\|^2\; > \;\|h\|^2 - (1-c_k^2)\|h\|^2\; =\; c_k^2\|h\|^2
\end{equation}
for every 
$h\in \rr{v}_0^\perp$. Hence $A_{P,\bot}:=P^\perp P_0^{\perp}|_{\rr{v}_0^\perp}:\rr{v}_0^\perp\to \rr{v}^\perp$ is an injective operator. Since $A_P|_{\rr{v}_0^\bot} =A_{P,\bot}$ to conclude the proof  the only missing point is the surjectivity, and in turn the invertibility, of $A_{P,\bot}$.
For that  it suffices to show that the positive operator $D_P:=A_{P,\bot}A_{P,\bot}^*:\rr{v}^\perp\to\rr{v}^\perp$ is invertible and therefore surjective. One has that
\[
D_P \;=\; P^\perp P_0^\perp P^\perp |_{\rr{v}^\perp} \;=\; (P^\perp - P^\perp P_0 P^\perp )|_{\rr{v}^\perp} \;=\; {\bf 1}_{\rr{v}^\perp}- P^\perp P_0 P^\perp|_{\rr{v}^\perp}\;,
\]
and since $\|P^\perp P_0 P^\perp|_{\rr{v}^\perp}\|\leq \|P^\perp P_0 \| < \sqrt{k}\varepsilon_k$, it follows that $D_P$ is invertible, hence surjective. 
Save for the weak continuity of $P\mapsto U_P = A_P|A_P|^{-1}$, we proved that $A_P = A_{P,0}+A_{P,\bot}$ provides the required  topological isomorphisms on the open set $P\in \s{O}_{0}$. To obtain the same result on the 
 neighbourhood $\s{O}_{Q}$ of a generic element $Q\in {^{\rm w}\bb{G}_k}(\rr{h})$
one can translate $P_0$, the neighbourhood $\s{O}_0$ and the section $s_0$ using any unitary operator $V$ such that $Q=VP_0V^*$. {To conclude the proof of the homeomorphism \eqref{eq:gra_set-w-fin} one can repeat verbatim
 the argument of \cite[Corollary 1.4]{shubin-96}.}

Finally, we prove the promised weak continuity of the map $\s{O}_0 \ni P\mapsto U_P = A_P |A_P|^{-1} \in {^{\rm w}}\bb{B}(\rr{h})$. For this we will first prove the weak continuity of $\s{O}_0 \ni P\mapsto |A_P|^{-1} = (A_P^*A_P)^{-1/2} \in {^{\rm w}}\bb{B}(\rr{h})$. We begin by observing that $A_P$ and
\[
A_P^*A_P \;=\; P_0 P P_0 + P_0^\perp P^\perp P_0^\perp\;,
\]
depend continuously on $P$ in the weak operator topology. We want to obtain a uniform positive lower bound on the spectrum of $A_P^* A_P$ in order to inherit the weak continuity to its inverse square root through functional calculus. 
Given that $\|A_{P,0}^{-1}\|<c_k^{-1}$ and $\|A_{P,\perp}^{-1}\| \leq c_k^{-1}$ (from equation \eqref{eq:proof_Rpz_lowerbound}) as shown above, it holds that for all $f\in\rr{h}$,
\[
\|A_P f \|^2 \;=\; \|A_P P f \|^2 + \|A_P P^\perp f \|^2 \;>\; c_k^2 \|P f \|^2 + c_k^2 \|P^\perp f \|^2 \;=\; c_k^2 \|f\|^2 \;.
\]
So $\|A_P f\| > c_k\|f\|$. Since $\|{(A_P^*)}^{-1}\| = \|{A_P}^{-1}\| \leq c_k^{-1}$, we also have $\|A_P^* f\| \geq c_k\|f\|$, and thus, for any $f\in\rr{h}$,
\[
\|A_P^*A_P f\| \;\geq\; c_k\|A_P f\| \;>\; c_k^2\|f\| \;.
\]
In particular, $\sigma(A_P^*A_P) \subseteq [c_k^2,1] \subsetneq[0,1]$.%, and $U_P = A_P |A_P|^{-1}$ is well-defined.

Now, let us prove inductively that $(A_P^*A_P)^m$ depends continuously (in the weak topology) on $P$ for any $m\in\N$. The case $m=1$ already holds. Assume it is true for all $m'=1,\dots,m-1$. Take a convergent net $P_\alpha\xrightarrow{\rm w} P$. Observe that by \cite[Proposition 3.1]{shubin-96}, this implies $P_\alpha\xrightarrow{\rm s} P$, where ${\rm s}$ denotes the strong topology, and thus $A_{P_\alpha}^*A_{P_\alpha} \xrightarrow{\rm s} A_P^*A_P$. Then, for all $f,g\in \rr{h}$
\begin{align*}
    \left\langle g, (A_{P_\alpha}^*A_{P_\alpha})^m f \right\rangle \;&=\; \left\langle (A_{P_\alpha}^*A_{P_\alpha})g, (A_{P_\alpha}^*A_{P_\alpha})^{m-1} f \right\rangle \\
    &\to\; \left\langle (A_P^*A_P)g, (A_P^*A_P)^{m-1} f \right\rangle \\
    &=\; \left\langle g, (A_P^*A_P)^m f \right\rangle \;,
\end{align*}
since $A_{P_\alpha}^*A_{P_\alpha} \xrightarrow{\rm s} A_P^*A_P$, $(A_{P_\alpha}^*A_{P_\alpha})^{m-1}\xrightarrow{\rm w} (A_P^*A_P)^{m-1}$, and both nets are uniformly bounded in norm. We deduce that $P\mapsto (A_P^*A_P)^m$ is weakly continuous for every $m\in\N$. With this at hand, it also holds that $P\mapsto p(A_P^*A_P)$ is weakly continuous for every polynomial $p$. Let $F:[c_k^2,1]\to\C$ be any continuous function. Fix $\delta>0$, and let $p_\delta$ be a polynomial such that $|F(t)-p_\delta(t)| < \delta/3$ for all $t\in [c_k^2,1]$. Then
\[
\| F(A_P^*A_P) - p_\delta(A_P^*A_P)\| \;=\; \sup_{t\in \sigma(A_P^*A_P)}|F(t)-p_\delta(t)| \;\leq\; \sup_{t\in [c_k^2,1]}|F(t)-p_\delta(t)| \;<\; \delta/3\;.
\]
Let $\alpha'$ be such that $\alpha\geq \alpha'$ implies $|\langle g,(p_\delta(A_P^*A_P)-p_\delta(A_{P_\alpha}^*A_{P_\alpha}))f\rangle| < \|g\|\|f\|\delta/3$. Then, by using the inequality above, it holds that for all $f,g\in \rr{h}$ and all $\alpha\geq\alpha'$,
\begin{align*}
    |\langle g,(F(A_{P_\alpha}^*A_{P_\alpha})-F(A_P^*A_P))f\rangle| 
    \;&\leq\; |\langle g,(F(A_{P_\alpha}^*A_{P_\alpha})-p_\delta(A_{P_\alpha}^*A_{P_\alpha}))f\rangle| \\
    &+\; |\langle g,(p_\delta(A_{P_\alpha}^*A_{P_\alpha})-p_\delta(A_P^*A_P))f\rangle| \\
    &+\; |\langle g,(p_\delta(A_P^*A_P)-F(A_P^*A_P))f\rangle| \\
    &<\; \|g\|\|f\| \delta \;.
\end{align*}
Hence, we conclude that $F(A_{P_\alpha}^*A_{P_\alpha})\xrightarrow{\rm w}F(A_P^*A_P)$, and thus $P\mapsto F(A_P^*A_P)$ is weakly continuous for any $F\in C([c_k^2,1],\C)$. In particular, taking $F(t) = t^{-1/2}$, we obtain that $P\mapsto |A_P|^{-1}$ is weakly continuous. Finally, we see that for all $f,g\in \rr{h}$,
\begin{align*}
    \left\langle g, U_{P_\alpha} f \right\rangle \;&=\; \left\langle A_{P_\alpha}^* g, |A_{P_\alpha}^*A_{P_\alpha}|^{-1} f \right\rangle \\
    &\to\; \left\langle A_P^*g, |A_P^*A_P|^{-1} f \right\rangle \\
    &=\; \left\langle g, U_P f \right\rangle \;,
\end{align*}
since again $A_{P_\alpha}^* \xrightarrow{\rm s} A_P^*$, $|A_{P_\alpha}^*A_{P_\alpha}|^{-1} \xrightarrow{\rm w} |A_P^*A_P|^{-1}$, and both nets are uniformly bounded in norm. We conclude that $U_{P_\alpha} \xrightarrow{\rm w} U_P$, and thus that $P\mapsto U_P$ is weakly continuous.
\end{proof}
% \qed

\medskip

The computation of the homotopy groups of ${^{\rm w}\bb{G}_n(\rr{h})}$
can be performed with the same argument used in \cite[Section 2.3]{abbondandolo-majer-09}. This consists in 
applying the  
long exact sequence  in homotopy associated to the (Serre) fibration
\[
\left[{^{\rm w}\bb{U}(\rr{v}_0)}\times {^{\rm w}\bb{U}(\rr{v}_0^\bot)}\right]\;\longrightarrow\;{^{\rm w}\bb{U}(\rr{h})}\;\stackrel{\pi}{\longrightarrow}\;{^{\rm w}\bb{G}_n(\rr{h})}
\]
obtained in Proposition \ref{prop:wheq}. By using the fact that ${^{\rm w}\bb{U}(\rr{h})}$ and ${^{\rm w}\bb{U}(\rr{v}_0^\bot)}$
are both contractible in view of the infinite dimensionality of 
$\rr{h}$ and $\rr{v}_0^\bot$ respectively, and that ${^{\rm w}\bb{U}(\rr{v}_0)}\simeq \n{U}(n)$ since $\rr{v}_0\simeq\C^n$, one obtains
\begin{corollary}
Let $n\in\N$.
The homotopy groups of ${^{\rm w}\bb{G}_n(\rr{h})}$ are given by
\[
\pi_k({^{\rm w}\bb{G}_n(\rr{h})})\;=\;\pi_k(B\n{U}(n))\;=\;\left\{
\begin{aligned}
&0&&\text{if}\;\; k=0&\\
&\pi_{k-1}(\n{U}(n))&&\text{if}\;\; k\geqslant 1\;.&\\
\end{aligned}
\right.
\]
% In particular, ${^{\rm w}\bb{G}_k}(\rr{h})$ is path-connected.
\end{corollary}

\medskip

Consider the bijective identification
\begin{equation}\label{eq:incl_col-w}
\imath\;:\;{^{\rm u}\bb{G}_n(\rr{h})}\;\to\; {^{\rm w}\bb{G}_n(\rr{h})}\;.
\end{equation}
This map is evidently continuous. 
\begin{proposition}
The map \eqref{eq:incl_col-w} provides a weak homotopy equivalence 
${^{\rm u}\bb{G}_n(\rr{h})} \thicksim{^{\rm w}\bb{G}_n(\rr{h})}$ for every $n\in\N_0$.
\end{proposition}
\proof
We already know that $\pi_k({^{\rm u}\bb{G}_n(\rr{h})})\simeq \pi_k({^{\rm w}\bb{G}_n(\rr{h})})$ for every $k\in\N_0$. Then,
one only needs to prove that the isomorphism is induced by $\imath_\ast$. For that let us simplify the notation. Let us define 
$F_u:= [{^{\rm u}\bb{U}(\rr{v}_0)}\times {^{\rm u}\bb{U}(\rr{v}_0^\bot)}]$, $U_u:={^{\rm u}\bb{U}(\rr{h})}$ and $G_u:={^{\rm u}\bb{G}_n(\rr{h})}$ and $F_w$, $U_w$ and $G_w$ the same spaces but with the weak topology. From \cite[Proposition 1.2]{shubin-96} and 
Proposition \ref{prop:wheq} we know that there are 
(Serre) fibrations (indeed locally trivial fiber bundles)
\[
F_u\;\stackrel{j'}{\longrightarrow}\;U_u\;\stackrel{\pi'}{\longrightarrow}\;G_u\;,\qquad F_w\;\stackrel{j}{\longrightarrow}\;U_w\;\stackrel{\pi}{\longrightarrow}\;G_w\;,
\]
where $j'$, $j$ are the inclusions of the fibers and $\pi'$, $\pi$ are the bundle projections. On top of that we have a commutative diagram
\[
\begin{tikzcd}
U_u \arrow[r,"\nu"] \arrow[d,swap,"\pi'"] & U_w  \arrow[d, "\pi"] \\
G_w \arrow[r,"\iota"]&   G_w 
\end{tikzcd}
\]
where the maps $\nu$ and $\iota$ are the bijective identifications which  correspond only to the change of topology of the spaces. 
These continuous maps induce homomorphisms in the associated long-exact sequences in homotopy
\[
\begin{tikzcd}
\cdots\arrow[r]  &\pi_k(U_u)\arrow[r,"\pi_*'"] \arrow[d,swap,"\nu_*"] & \pi_k(G_u)  \arrow[d,swap, "\iota_*"] \arrow[r,]& \pi_{k-1}(F_u)\arrow[r,"j_*'"]\arrow[d,swap,"f_*"] \arrow[r]&\pi_{k-1}(U_u)\arrow[d,swap,"\nu_*"]\arrow[r]&\cdots\\
\cdots \arrow[r]& \pi_k(U_w) \arrow[r,"\pi_*"]&  
 \pi_k(G_w) \arrow[r] &\pi_{k-1}(F_w)\arrow[r,"j_*"] \arrow[r]&\pi_{k-1}(U_w)\arrow[r]&\cdots
\end{tikzcd}
\]
where the vertical arrow $f_*$ is the homomorphism induced by the continuous map $f:F_u\to F_w$ obtained as the restriction of the map $\nu:U_u\to U_w$ on the fiber. Again $f$ is the bijective identification which  corresponds only to the change of topology of the corresponding spaces.
Since both $U_u$ and $U_w$ are contractible spaces one gets
\[
\begin{tikzcd}
0\arrow[r,"\pi_*'"]  & \pi_k(G_u)  \arrow[d,swap, "\iota_*"] \arrow[r,]& \pi_{k-1}(F_u)\arrow[r,"j_*'"]\arrow[d,swap,"f_*"] \arrow[r]&0\\
 0 \arrow[r,"\pi_*"]&  
 \pi_k(G_w) \arrow[r] &\pi_{k-1}(F_w)\arrow[r,"j_*"] \arrow[r,]&0
\end{tikzcd}
\]
which provides that $\iota_\ast:\pi_k(G_u)\to \pi_k(G_w)$ is an isomorphism.
\qed

\medskip

By the transitivity of weak homotopy equivalence, one gets:

\begin{corollary}
There is a weak homotopy equivalence $B\n{U}(n) \thicksim {^{\rm w}\bb{G}_n(\rr{h})}$ for all $n\in\N_0$.
\end{corollary}

%--------%
\section{The indefinite rank case}\label{sec:infin-R}
Let us introduce the spaces 
\begin{equation}\label{eq:gra_1k-<N}
{^\sharp\bb{G}_{\bullet,\le N}(\rr{h})}\;:=\;\bigcup_{n=0}^N{^\sharp\bb{G}_n(\rr{h})}\; =\;{^\sharp\{P\in\bb{P}(\rr{h})\;|\; {\rm Tr}_{\rr{h}}(P)\leqslant N\}}
\end{equation}
where $\sharp=\{{\rm u},{\rm w}\}$ denotes the possible topology.
One has evident inclusions ${^\sharp\bb{G}_{\bullet,\le N}(\rr{h})}\subset {^\sharp\bb{G}_{\bullet,{\le N+1}}(\rr{h})}$ and in turn
\begin{equation}\label{eq:gra_1k-<N>inf}
{^\sharp\bb{G}_{\bullet}(\rr{h})}\;=\;\bigcup_{N=0}^\infty{^\sharp\bb{G}_{\bullet,\le N}(\rr{h})}
\end{equation}
has the set-theoretical structure of an inductive limit.

\medskip

Let us observe that each ${^{\rm u}\bb{G}_n(\rr{h})}$ is closed in  $\bb{B}(\rr{h})$ \cite[Lemma 1.1]{shubin-96}. In fact  ${^{\rm u}\bb{G}_n(\rr{h})}$ is a complete metric space with respect to the operator norm. This means that 
\begin{equation}\label{eq:gra_1k-<N>inf??}
{{^{\rm u}\bb{G}_{\bullet}(\rr{h})}}\;=\;\bigsqcup_{n=0}^\infty{^{\rm u}\bb{G}_{n}(\rr{h})}
\end{equation}
is indeed the disjoint union of closed spaces each of which has the homotopy type
of a distinct classifying space $B\n{U}(n)\thicksim{^{\rm u}\bb{G}_n(\rr{h})}$ (see Appendix \ref{unif_gras}). This observation paves the way for a first no-go result.
\begin{lemma}\label{lemm_nogo}
The spaces ${{^{\rm u}\bb{G}_{\bullet}(\rr{h})}}$ and $B\n{U}(\infty)$ have different homotopy type.
\end{lemma}
\proof
We know that each ${^{\rm u}\bb{G}_{n}(\rr{h})}$ is a path connected closed subspace of ${{^{\rm u}\bb{G}_{\bullet}(\rr{h})}}$. Note that if $P_0\in {^{\rm u}\bb{G}_{n}(\rr{h})}$, then $\{P\in {{^{\rm u}\bb{G}_{\bullet}(\rr{h})}} \;|\; \|P-P_0\|<1\} \subseteq {^{\rm u}\bb{G}_{n}(\rr{h})}$. Since $P_0$ and $n\in\N_0$ were arbitrary, it follows that each ${^{\rm u}\bb{G}_{n}(\rr{h})}$ is open, and hence a connected component of ${{^{\rm u}\bb{G}_{\bullet}(\rr{h})}}$. Therefore, the ${^{\rm u}\bb{G}_{n}(\rr{h})}$ are path components of ${{^{\rm u}\bb{G}_{\bullet}(\rr{h})}}$, and together with Equation \eqref{eq:gra_1k-<N>inf??}, one can see that $\pi_0({{^{\rm u}\bb{G}_{\bullet}(\rr{h})}})=\N_0$. On the other hand $\pi_0(B\n{U}(\infty))=0$, since this space is path connected. Additionally, after fixing the base point $\ast$ in one of the components of ${{^{\rm u}\bb{G}_{\bullet}(\rr{h})}}$, say for instance $\ast\in{^{\rm u}\bb{G}_{\bullet,n}(\rr{h})}$, then 
$\pi_k({{^{\rm u}\bb{G}_{\bullet}(\rr{h})}})=\pi_k(B\n{U}(n))$
(with respect to the given base point). However, the series of homotopy groups of the spaces $B\n{U}(n)$ is different from the series of homotopy groups of  $B\n{U}(\infty)$. \qed

\medskip

Let us explore now the case of the weak topology. The first relevant observation is  \cite[Proposition 3.4 (ii)]{shubin-96} which shows that
\[
\overline{{^{\rm w}\bb{G}_{n}(\rr{h})}}\;=\;{^{\rm w}\bb{G}_{\bullet,\le n}(\rr{h})}
\]
for every $n\in \N$, where the closure is meant with respect to the weak-topology \emph{inside} the projections. In particular this fact implies that the spaces  ${^{\rm w}\bb{G}_{n}(\rr{h})}$ are not disconnected as in the case of the uniform topology, and for that we are not in the situation of Lemma \ref{lemm_nogo}. From the equality above one also gets
\[
\overline{^{\rm w}\bb{G}_{\bullet,\le N}(\rr{h})}\;=\;{^{\rm w}\bb{G}_{\bullet,\le N}(\rr{h})}
\]
for every $N\in \N$, showing that these spaces are indeed closed in the weak topology inside the projections. The homotopy properties of these spaces are described in the following result. 
\begin{proposition}\label{prop_nogo-2}
The space ${^{\rm w}\bb{G}_{\bullet}(\rr{h})}$ is  contractible. 
The spaces ${^{\rm w}\bb{G}_{\bullet,\le N}(\rr{h})}$ are contractible for any $N\in\N_0$. 
\end{proposition}
\proof
Since $\rr{h}$ is separable there is a (not unique) unitary map
$U:\rr{h}\to L^2([0,1])$. Since a unitary map preserves the property of being a projection and the trace one gets that $\bb{G}_{\bullet}(\rr{h})=U\bb{G}_{\bullet}U^{-1}$ and $\bb{G}_{\bullet,\le N}(\rr{h})=U\bb{G}_{\bullet,\le N}U^{-1}$, where the sets 
$\bb{G}_{\bullet}$ and $\bb{G}_{\bullet,\le N}$ have been introduced in Appendix \ref{weak_contr}. The unitary equivalence also preserves the weak topology, implying that ${^{\rm w}\bb{G}_{\bullet}(\rr{h})}$ and 
${^{\rm w}\bb{G}_{\bullet,\le N}(\rr{h})}$ have the same topological properties of ${^{\rm w}\bb{G}_{\bullet}}$ and 
${^{\rm w}\bb{G}_{\bullet,\le N}}$ respectively. The rest of the proof follows from Proposition \ref{prop_def_ref} and Remark 
\ref{rk_def}.
\qed

\medskip

The result above implies our second no-go result.
\begin{corollary}\label{cor_nogo}
The spaces ${{^{\rm w}\bb{G}_{\bullet}(\rr{h})}}$ and $B\n{U}(\infty)$ have different homotopy type.
\end{corollary}
%

%---%
\section{Continuous path of states}\label{sec:pat_sta}
This section is devoted to the proof of Proposition \ref{prop:con_stat}.

\medskip

Observe 
since ${\rm Tr}_{\rr{h}}(P(x))\in\N$ the continuity of 
$x\mapsto {\rm Tr}_{\rr{h}}(P(x))$ implies that ${\rm Tr}_{\rr{h}}(P(x))$ must have constant values on the connected components of $X$.
Then, without loss of generality, we can assume that $X$ is connected and ${\rm Tr}_{\rr{h}}(P(x))=M$ for every $x\in X$. On the contrary one can repeat the argument on each connected component of $X$.

\medskip

Let $x_0\in X$ be a given point and $U\in\bb{A}$ a unitary element.
Then
\[
\begin{aligned}
\big|\omega_x(U)-\omega_{x_0}(U)\big|\;&=\;\frac{1}{M}\big|{\rm Tr}_{\rr{h}}(P(x)U)-{\rm Tr}_{\rr{h}}(P(x_0U))\big|\\
\;&\leqslant\;\|P(x)U-P(x_0)U\|_1\\
&=\;\|Q_x-Q_{x_0}\|_1
\end{aligned}
\] 
where $\|\cdot\|_1$ denotes the norm of the Schatten ideal $\bb{L}^1(\rr{h})$ of trace class operators and $Q_x:=P(x)U$ for every $x\in X$. One has that $Q_x\to Q_{x_0}$  in the weak topology.
Moreover $|Q_x|=U^*P(x)U$ and $|Q_x^*|=P(x)$ for every $x\in X$
showing that also $|Q_x|\to|Q_{x_0}|$ and $|Q_x^*|\to|Q_{x_0}^*|$ in the weak topology. Finally $\|Q_x\|_1={\rm Tr}_{\rr{h}}(U^*P(x)U)=M$
for every $x\in X$. Since all the condition of \cite[Theorem 2.20]{simon-05} are verified one gets that $\|Q_x-Q_{x_0}\|_1\to 0$ as $x\to x_0$, and in turn $\omega_x(U)\to \omega_{x_0}(U)$ for every unitary $U$. To complete the proof it is enough to recall that each $A\in\bb{A}$ is the sum of  at most four unitaries $A=\sum_{j=1}^4 a_jU_j$ with $a_j\in\C$ bounded by $|a_j|\leqslant \|A\|/2$
\cite[Lemma 2.2.14]{bratteli-robinson-87}, and to use the linearity of $\omega_x$.
\qed

\appendix

%------------------------------------------------------------------------%
\section{Classical Grassmannians}
\label{sec:class_Grass}
The classifying spaces $B\n{U}(n)$ can be modeled out of the classical \emph{Grassmannian} as described in 
\cite[Section 1.2]{hatcher-17} and \cite[Chapters 5 \& 6]{milnor-stasheff-74}.
For non negative integers $n \leqslant d$, let 
\begin{equation}\label{eq:sym_sp}
{\rm Gr}_{n}(\C^d)\;\simeq\;\n{U}(d)/[\n{U}(n)\times\n{U}(d-n)]
\end{equation}
be the set of all the possible subspaces of dimension $n$ inside $\C^d$ or, equivalently,  the set of all  possible $n$-planes of $\C^d$ containing the origin. For  $k=1$ one has that ${\rm Gr}_{1}(\C^d)=\n{P}(\C^d)$ coincides with the projective space of $\C^d$. 
Any ${\rm Gr}_{n}(\C^d)$ can be endowed with the quotient topology making it into a Hausdorff  and path-connected manifold with the structure of a finite CW-complex.
The inclusions $\C^d\subset\C^{d+1}\subset\ldots$ obtained 
identifying the vector $v \in \C^d$ with the vector  $(v,0)\in \C^{d+1}$ yield inclusions ${\rm Gr}_{n}(\C^d)\subset {\rm Gr}_{n}(\C^{d+1})\subset\ldots$ and one can consider the union
\begin{equation}\label{eq:gra_0k}
{\rm Gr}_{n,\infty}\;:=\; \bigcup_{d=n}^\infty {\rm Gr}_{n}(\C^d)\;.
\end{equation}
The space ${\rm Gr}_{n,\infty}$ consists of all the possible $n$-plane of $\C^\infty$ containing the origin.
Given the injections $\imath_d:{\rm Gr}_{n}(\C^d)\hookrightarrow {\rm Gr}_{n,\infty}$ one can endow ${\rm Gr}_{n,\infty}$ with the \emph{inductive limit topology}, that is the final topology given by  the set of the functions ${\imath_d}$.  The space ${\rm Gr}_{n,\infty}$ inherits a $CW$-complex structure and it results to be a Hausdorff, paracompact and path-connected space.
A relevant case is $n=1$ for which ${\rm Gr}_{1,\infty}=\n{P}(\C^\infty)\sim K(\Z,2)$, where $\n{P}(\C^\infty)$ is the projective space of $\C^\infty$ and $K(\Z,2)$ is the  Eilenberg-MacLane space.
In this case one obtains that 
\begin{equation}\label{eq_class_vect_classic2}
[X,{\rm Gr}_{1,\infty}]\;\simeq\;[X,K(\Z,2)]\;\stackrel{c_1}{\simeq}\;H^2(X,\Z)
\end{equation}
where the map $c_1$ which provides the bijection (indeed a group homomorphism) is known as \emph{first Chern class}.

\medskip

For the infinite Grassmannian 
the inclusions   ${\rm Gr}_{n}(\C^d)\subset {\rm Gr}_{n+1}(\C^{d+1})\subset\ldots$, which consist of adding the subspace generated by the last vector $e_{d+1}:=(0,\ldots,0,1)\in\C^{d+1}$, induce injections $\jmath_n:{\rm Gr}_{n,\infty}\hookrightarrow {\rm Gr}_{n+1,\infty}$. One can define the space of all possible $n$-planes of $\C^\infty$ with any $n$ containing the origin as
\begin{equation}\label{eq:gra_0-inf}
{\rm Gr}_{\infty}\;:=\; \bigcup_{n=0}^\infty {\rm Gr}_{n,\infty}
\end{equation}
endowed with the {inductive limit topology} induced by the maps $\jmath_k$. It turns out that ${\rm Gr}_{\infty}$ is again a Hausdorff, paracompact and path-connected space.

%------------------------------------------------------------------------%
\section{Infinite unitary group}\label{app_unit_op}
Let $\rr{h}$ be a separable Hilbert space with
 ${\rm dim}(\rr{h})=\aleph_0$ 
and 
$\bb{U}(\rr{h})\subset \bb{B}(\rr{h})$ the group of unitary operators.
One can induce on $\bb{U}(\rr{h})$ any of the topologies supported on $\bb{B}(\rr{h})$.
 There are at least three (indeed more) relevant topologies: the norm (or uniform) topology, the strong topology and the weak topology. As shown in \cite{espinoza-uribe-14}
there are only two relevant cases: (1) Let $^{\rm u}\bb{U}(\rr{h})$ be $\bb{U}(\rr{h})$ endowed with the uniform topology. Then  $^{\rm u}\bb{U}(\rr{h})$
 is a Banach-Lie group. (2) All the other topologies on $\bb{U}(\rr{h})$ coincide, and we shall use the symbol $^{\rm w}\bb{U}(\rr{h})$ to denote $\bb{U}(\rr{h})$ endowed with the weak topology just to fix one. Then $^{\rm w}\bb{U}(\rr{h})$ is a Polish group. In particular it is a completely metrizable space.
It is known that both 
$^{\rm u}\bb{U}(\rr{h})$ and $^{\rm w}\bb{U}(\rr{h})$ are  contractible \cite{kuiper-65,schottenloher-18}. Therefore the 
 bijective identification
\begin{equation}\label{eq:incl_col-w-unit}
\imath\;:\;{^{\rm u}\bb{U}(\rr{h})}\;\hookrightarrow\; {^{\rm w}\bb{U}(\rr{h})}\;
\end{equation}
provides a homotopy equivalence ${^{\rm u}\bb{U}(\rr{h})}\simeq {^{\rm w}\bb{U}(\rr{h})}$. Considering the inclusions $\n{U}(n)\subset\n{U}(n+1)$ and the inductive limit
\begin{equation}\label{eq:gra_0-inf-unitrars}
\n{U}(\infty)\;:=\; \bigcup_{n=1}^\infty \n{U}(n)
\end{equation}
one knows that $\pi_k(\n{U}(\infty))=0$ if $k$ is $0$ or even, and 
$\pi_k(\n{U}(\infty))=\Z$ if $k$ is odd (Bott periodicity). Therefore, 
$\n{U}(\infty)$ cannot be weakly homotopy equivalent to ${^{\rm u}\bb{U}(\rr{h})}$ or ${^{\rm w}\bb{U}(\rr{h})}$.
In fact $\n{U}(\infty)$ has the same homotopy type of the subgroup  ${^{\rm u}\bb{U}_{c}(\rr{h})}\subset {^{\rm u}\bb{U}(\rr{h})}$ made of unitary operators $U$ such that $U-{\bf 1}\in\bb{K}(\rr{h})$ is compact \cite{palais-65}.

\section{The uniform Grassmannian as classifying space}\label{unif_gras}
In this section we will give an explicit proof of the fact that the map \eqref{eq:incl_col} provides  a weak homotopy equivalence
${\rm Gr}_{n,\infty}\sim {^{\rm u}\bb{G}_n(\rr{h})}$.

\medskip

 Taking a complete orthonormal basis $\{ e_n\}_{n\in\N}$ of $\mathfrak{h}$, we have the identifications 
 $\C^d\simeq\mathfrak{h}_d := \mathrm{span}(e_1, \ldots, e_d)\subset\mathfrak{h}$ for every $d\in\N$. This identification 
  induces a homeomorphism $\mathrm{Gr}_n(\C^d) \simeq \mathscr{G}_n(\mathfrak{h}_d)$. For an orthogonal projection $P \in \mathrm{Gr}_n(\C^d)$ (under the aforementioned identification),
one gets  $P \oplus {\bf 0}_{\mathfrak{h}_d^\perp} \in \mathscr{G}_n(\mathfrak{h})$, where ${\bf 0}_{\mathfrak{h}_d^\perp}$ is the null-map on the orthogonal complement $\mathfrak{h}_d^\perp$ of $\mathfrak{h}_d$ in $\mathfrak{h}$. Then we get a continuous injection
$$
\jmath_d\; :\; \mathrm{Gr}_n(\C^d) \;\hookrightarrow\; {^{\rm u}\bb{G}_n(\rr{h})}$$
given by $P \mapsto P \oplus {\bf 0}_{\mathfrak{h}_d^\perp}$. This injection is compatible with the linear embedding $\C^d \subset \C^{d+1}$, so that the $\jmath_d$'s together define a continuous map
$$
\jmath \;:\; \mathrm{Gr}_n(\C^\infty) \;\rightarrow\; {^{\rm u}\bb{G}_n(\rr{h})}\;.
$$

\medskip

To prove that $\jmath$ is a weak homotopy equivalence we will make use of the homotopy exact sequence. Let us consider the classical homeomorphism
\eqref{eq:sym_sp} which represents $\mathrm{Gr}_n(\C^d)$ as a homogeneous space, and the equivalent homeomorphism \eqref{eq:gra_set-u} for ${^{\rm u}\bb{G}_n(\rr{h})}$ with the choice of 
$\mathfrak{v}_0 = \mathfrak{h}_n$.
We have a continuous monomorphism $\n{U}(d) \simeq {^{\rm u}\bb{U}(\mathfrak{h}_d)} \to {^{\rm u}\bb{U}(\rr{h})}$ given by $U \mapsto U \oplus {\bf 1}_{\mathfrak{h}_d^\perp}$, where ${\bf 1}_{\mathfrak{h}_d^\perp}$ is the identity on the orthogonal complement $\mathfrak{h}_d^\perp$ of $\mathfrak{h}_d$ in $\mathfrak{h}$. The induced map on the quotient spaces turns out to be $\jmath_d$. 
Thus, we get a map of principal bundles %
\[
\begin{tikzcd}
\n{U}(d)/\n{U}(d-n)\arrow[r,"\widetilde{\jmath}_d"] \arrow[d] & {^{\rm u}\bb{U}(\rr{h})}/{^{\rm u}\bb{U}(\rr{h}_n^{\bot})}   \arrow[d] \\
\mathrm{Gr}_n(\C^d) \arrow[r,swap,"\jmath_d"]&   {^{\rm u}\bb{G}_n(\rr{h})}
\end{tikzcd}
\]
whose structure groups are $\n{U}(n) \simeq {^{\rm u}\bb{U}(\rr{h}_n)}$. Taking  the direct limit, one also gets a map of principal bundles (with the same structure group)
\[
\begin{tikzcd}
\mathrm{S}_{n,\infty}\arrow[r,"\widetilde{\jmath}"] \arrow[d] & {^{\rm u}\bb{U}(\rr{h})}/{^{\rm u}\bb{U}(\rr{h}_n^{\bot})}   \arrow[d] \\
\mathrm{Gr}_{n,\infty} \arrow[r,swap,"\jmath"]&   {^{\rm u}\bb{G}_n(\rr{h})}
\end{tikzcd}
\]
where $\mathrm{S}_{n,\infty}:=\lim_{d\to \infty}\n{U}(d)/\n{U}(d-n)$.
It follows a homomorphism of homotopy exact sequences:
\[
\begin{tikzcd}%[row sep=huge]
\vdots  \arrow[d]&\vdots  \arrow[d]\\
\pi_k(\n{U}(n))\arrow[r,swap,"{\jmath}_\ast"] \arrow[d] & \pi_k({^{\rm u}\bb{U}(\rr{h}_n)})   \arrow[d] \\
\pi_k(\mathrm{S}_{n,\infty}) \arrow[r,swap,"\widetilde{\jmath}_\ast"]
\arrow[d]&  \pi_k({^{\rm u}\bb{U}(\rr{h})}/{^{\rm u}\bb{U}(\rr{h}_n^{\bot})} )\arrow[d]\\
\pi_k(\mathrm{Gr}_{n,\infty}) \arrow[r,swap,"{\jmath}_\ast"]   \arrow[d]&\pi_k({^{\rm u}\bb{G}_n(\rr{h})}) \arrow[d]\\
\pi_{k-1}(\n{U}(n))\arrow[r,swap,"{\jmath}_\ast"]   \arrow[d] & \pi_{k-1}({^{\rm u}\bb{U}(\rr{h}_n)})   \arrow[d] \\
\vdots&\vdots
\end{tikzcd}
\]
The injection $\jmath$ induces a homeomorphic homomorphism $\jmath : \n{U}(n) \to {^{\rm u}\bb{U}(\rr{h}_n)}$. Hence one gets isomorphisms $\pi_k(U(\C^n)) \simeq \pi_k(\mathscr{U}(\mathfrak{v}_0))$ for all $k$
induced by the maps $\jmath_*$.
 Now, to complete the proof it is enough to show that
$$
\widetilde{\jmath}_\ast\; : \;
\pi_k(\mathrm{S}_{n,\infty})
\;\longrightarrow\;
\pi_k({^{\rm u}\mathscr{U}(\mathfrak{h}})/{^{\rm u}\mathscr{U}(\rr{h}_n^\perp)})
$$
is an isomorphism for all $k$. This simply follows from the vanishing of these  homotopy groups. In fact $\n{U}(d)/\n{U}(d-n)$
 is nothing but the Stiefel manifold. It is known \cite[Theorem 5.1]{husemoller-94} that $\pi_k(\n{U}(d)/\n{U}(d-n)) = 0$ for $k \leqslant 2n$. By applying \cite[Proposition 4.3]{husemoller-94}
 one obtains $\pi_k(\mathrm{S}_{n,\infty})=0$  for all $k$.
According to \cite[Proposition 1.2]{shubin-96}, the projection ${^{\rm u}\mathscr{U}(\mathfrak{h})} \to {^{\rm u}\mathscr{U}(\mathfrak{h})}/[{^{\rm u}\mathscr{U}(\rr{h}_n)} \times {^{\rm u}\mathscr{U}(\rr{h}_n^\perp)}]$ admits a local section. Using this local section, one can get a local section of the projection ${^{\rm u}\mathscr{U}(\mathfrak{h})} \to {^{\rm u}\mathscr{U}(\mathfrak{h})}/{^{\rm u}\mathscr{U}(\rr{h}_n^\perp)}$. Thus, the latter is a principal ${^{\rm u}\mathscr{U}(\rr{h}_n^\perp)}$-bundle. Since the homotopy groups of ${^{\rm u}\mathscr{U}(\mathfrak{h})}$ and ${^{\rm u}\mathscr{U}(\rr{h}_n^\perp)}$ are trivial by Kuiper's theorem \cite{kuiper-65}, so are the homotopy groups of  ${^{\rm u}\mathscr{U}(\mathfrak{h})}/{^{\rm u}\mathscr{U}(\rr{h}_n^\perp)}$ by the homotopy exact sequence.
In conclusion, one obtains isomorphisms $\jmath_*:\pi_k(\mathrm{Gr}_{n,\infty})\to \pi_k({^{\rm u}\bb{G}_n(\rr{h})})$ for every $k$. 
This implies that $\jmath$ is a weak homotopy equivalence.

\section{Weak contraction of projections}\label{weak_contr}
This section contains some technical results inspired by \cite[Appendix 2]{Atiyah-segal-04}.

\medskip

Let $L^2([0, 1])$ be the separable Hilbert space of the $L^2$-functions on the unit interval. For each $t \in [0, 1]$, consider the  orthogonal decomposition
\[
L^2([0, 1]) \;=\; L^2([0, t]) \oplus L^2([t, 1])\;.
\]
Let $R_t : L^2([0, 1]) \to L^2([0, 1])$ be the orthogonal projection onto $L^2([0, t])$. More precisely, if $f\in L^2([0, 1])$ then
$(R_tf)(x):=\chi_{[0,t]}(x)f(x)$ where $\chi_I$ denotes the characteristic function of the set $I\subseteq [0,1]$. 
Let us introduce the domain restriction projection $\pi_t$ and the inclusion $\iota_t$, defined as maps
\begin{align}
\pi_t &\;:\; L^2([0, 1])\; \longrightarrow\; L^2([0, t]), &
\iota_t &\;:\; L^2([0, t])\; \longrightarrow\; L^2([0, 1]),
\end{align}
As operations on an element of $L^2([0, 1])$ the projection $\pi_t$ consists in the restriction of the domain, and $\iota_t$ is the extension of the domain by zero. One has the  decomposition $R_t = \iota_t  \pi_t$. Finally, for $0 < t \le 1$, let  $\sigma_t : L^2([0, t]) \to L^2([0, 1])$ be the isometric isomorphism defined
by
\begin{align}
(\sigma_t f)(x) &\;:=\; \sqrt{t}\;f(tx)\;.
\end{align}
Let $\bb{B}:=\bb{B}(L^2([0, 1]))$ be the set of bounded operators on $L^2([0, 1])$. For every $0 < t \le 1$ consider the map
$\phi_t:\bb{B}\to \bb{B}$ defined by
\begin{align}
\phi_t(A) &\;:=\; \iota_t \sigma_t^{-1}A\sigma_t \pi_t\;,\qquad A\in\bb{B}\;.
\end{align}
From the definition above, one immediately infers that $\phi_1(A)=A$ for every $A\in\bb{B}$.
\begin{lemma}\label{lemm_appD1}
For every $0 < t \le 1$ the map $\phi_t$ is linear.
Moreover,
\[
\phi_t(A^*)\;=\;\phi_t(A)^*\;,\qquad \phi_t(AB)\;=\;\phi_t(A)\phi_t(B)
\]
for every $A,B\in\bb{B}$.
\end{lemma}
\proof
The linearity is evident. Let $D_t:=\sigma_t \pi_t\in\bb{B}$.
By observing that $\iota_t$ and $\pi_t$ are adjoint to each other, and $\sigma_t^*=\sigma_t^{-1}$ since it is an isometry, one gets that
$D_t^*=\iota_t \sigma_t^{-1}$. This means that $\phi_t(A)=D_t^*AD_t$
and therefore the adjoint property follows immediately.
For the 
 product property let us observe that  $\pi_t\iota_t$ is the identity on $L^2([0, t])$. Therefore $D_tD_t^*={\bf 1}$ is the identity  on $L^2([0, 1])$ and $D_t^*D_t=R_t$ is the subspace projection introduced above. The first one of these relations is enough to show the product formula. 
\qed

\medskip

Let $\mathscr{G}_{\bullet}$ be the set of trace-class orthogonal projections of $L^2([0, 1])$.
\begin{corollary}
For every $0 < t \le 1$, $\phi_t$ restricts to a well-defined map $\phi_t:\mathscr{G}_{\bullet}\to \mathscr{G}_{\bullet}$.
\end{corollary}
\proof
Let $P$ be an orthogonal projection. Then, in view of Lemma \ref{lemm_appD1} it follows that $\phi_t(P)^*=\phi_t(P^*)=\phi_t(P)$
and $\phi_t(P)^2=\phi_t(P^2)=\phi_t(P)$. Therefore $\phi_t(P)$ is a projection. Moreover
\[
{\rm Tr}\big(\phi_t(P)\big)\;=\;{\rm Tr}\left(D_t^*PD_t\right)\;\leqslant\;{\rm Tr}\left(P\right)
\]
since $D_t$ is a partial isometry, and in turn $\|D_t\|\leqslant 1$.
\qed

\medskip

Let $\bb{G}_{\bullet,\le N}$ be the set of orthogonal projections of $L^2([0, 1])$ of rank at most $N$. Since $\phi_t$ doesn't increase the trace one obtains the more precise result:
\begin{corollary}\label{cor_insid_N}
For every $0 < t \le 1$ and every $N\in\N_0 $, it holds that $\phi_t:\bb{G}_{\bullet,\le N}\to \bb{G}_{\bullet,\le N}$.
\end{corollary}

\medskip

We are now interested in the behaviour of $\phi_t$ when $t\to 0$.
\begin{lemma}\label{lemm_lim_0}
For every $A\in\bb{B}$ one has that $\phi_t(A)\to0$ when $t\to 0$ with respect to the weak topology.
\end{lemma}
\proof
Let $f\in L^2([0, 1])$ be a normalized function. From the identity 
\[
\|D_tf\|^2\;=\; \int_0^1 \dd x \; t|f(tx)|^2 \;=\; \int_0^t\dd x\;|f(x)|^2
\]
where $D_t$ is defined as above, one deduces that $\|D_t f\|\to 0$ when $t\to 0$. 
Therefore from
\[
|\langle f,\phi_t(A) g\rangle|\;\leqslant\;\|A\|\;\|D_tf\| \|D_tg\|
\]
one gets that $\langle f,\phi_t(A) g\rangle \to 0$
for every $f,g\in L^2([0, 1])$ when $t\to 0$.
\qed

\medskip

In view of the above, we can define $\phi_0(A):=0$, so for each $A\in \bb{B}$, $t\mapsto \phi_t(A)$ is continuous in the weak topology as a function $[0,1]\to \bb{B}$. Let us denote with ${^{\rm w}\bb{G}_{\bullet,\le N}}$ the space of trace-class orthogonal projections endowed with the weak topology.
We are in position to prove the main result of this section.
\begin{proposition}\label{prop_def_ref}
Let $\Phi:[0,1]\times {^{\rm w}\bb{G}_{\bullet}}\to {^{\rm w}\bb{G}_{\bullet}}$ be the map defined by $\Phi(t,P):=\phi_{1-t}(P)$. Then $\Phi$ is a continuous map such that $\Phi(0,P)=P$ and $\Phi(1,P)=0$ for every $P\in {^{\rm w}\bb{G}_{\bullet}}$.
\end{proposition}
\proof
For every $t,t'\in[0,1]$ and $P,P'\in {^{\rm w}\bb{G}_{\bullet}}$ one has that
\begin{align*}
    \Phi(t',P')-\Phi(t,P)\;&=\; D_{1-t'}^* P' D_{1-t'} - D_{1-t}^* P D_{1-t} \\
    &=\; D_{1-t'}^* P' (D_{1-t'} - D_{1-t}) \\
    &+\; (D_{1-t'}^* - D_{1-t}^*)P'D_{1-t} \\
    &+\; D_{1-t}^* (P'-P) D_{1-t} \;,
\end{align*}
where we just added and subtracted $D_{1-t'}^* P' D_{1-t}$ and $D_{1-t}^* P' D_{1-t}$.
Therefore, in order to prove that
\[
\lim_{(t',P')\to(t,P)}\big|\langle f,(\Phi(t',P')-\Phi(t,P))g \rangle\big|\;=\;0
\]
for every $f,g\in L^2([0, 1])$, it is sufficient to prove that
\begin{align}
    &\lim_{(t',P')\to(t,P)}\big|\langle f,(D_{1-t'}^* P' (D_{1-t'} - D_{1-t}))g \rangle\big|\;=\;0 \label{eq:lim_001} \;,\\
    &\lim_{(t',P')\to(t,P)}\big|\langle f,((D_{1-t'}^* - D_{1-t}^*)P'D_{1-t})g \rangle\big|\;=\;0 \label{eq:lim_002} \;, \text{ and}\\
    &\lim_{(t',P')\to(t,P)}\big|\langle f,(D_{1-t}^* (P'-P) D_{1-t})g \rangle\big|\;=\;0 \;.\label{eq:lim_003}
\end{align}
Let us assume for the moment the continuity of the map $s\mapsto D_s$  with respect to the strong topology. We can use this to prove Equation \eqref{eq:lim_001} as follows:
\begin{align*}
    \big|\langle f,(D_{1-t'}^* P' (D_{1-t'} - D_{1-t}))g \rangle\big| \;&=\; \big|\langle P' D_{1-t'}f,(D_{1-t'} - D_{1-t})g \rangle\big| \\
    &\leq\; \|P' D_{1-t'}f\| \|(D_{1-t'} - D_{1-t})g\| \\
    &\leq\; \|f\|\|(D_{1-t'} - D_{1-t})g\| \xrightarrow{(t',P')\to(t,P)} 0 \;.
\end{align*}
By the same argument, Equation \eqref{eq:lim_002} holds true. Finally, observe that
\[
    |\langle f,(D_{1-t}^* (P'-P) D_{1-t})g \rangle\big| \;=\; |\langle D_{1-t}f, (P'-P) D_{1-t}g \rangle\big| \xrightarrow{(t',P')\to(t,P)} 0 \;,
\]
due to the weak convergence of $P'\to P$, so Equation \eqref{eq:lim_003} holds true as well. Therefore, the Proposition follows if we prove the strong continuity of $s\mapsto D_s$. 
We already know it to be continuous at $s=0$.

To conclude the argument let us observe that for $f\in L^2([0, 1])$ and $0 < s \le 1$ one has that
\[
(D_sf)(x)\;=\;\sqrt{s}\; f|_s(sx)
\]
where $f|_s:=\pi_s f$ denotes the restriction of $f$ on $[0,s]$.
Therefore, for any $s'\in(0,1]$,
\[
\begin{aligned}
\|(D_{s'}-D_s)f\|^2\;&=\;\int_0^1\dd x\;\left|\sqrt{s'}\; f|_{s'}(s'x)-\sqrt{s}\; f|_s(sx)\right|^2\;\leqslant\;a_f(s',s)+b_f(s',s)\\
\end{aligned}
\]
where
\[
a_f(s',s)\;:=\;2\left|\sqrt{s'}-\sqrt{s}\right|^2\int_0^1\dd x\;\; \big|f|_{s'}(s'x)\big|^2\;\leqslant\; 2 \frac{\left|\sqrt{s'}-\sqrt{s}\right|^2}{s'}\|f\|^2
\]
and
\[
b_f(s',s)\;:=\;2s\int_0^1\dd x\;\big|  f|_{s'}(s'x)- f|_s(sx)\big|^2\;.
\]
From the inequality which dominates $a_f(s',s)$ it follows that $a_f(s',s)\to 0$ when $s'\to s$ for every $f\in L^2([0, 1])$.
For the term $b_f(s',s)$ let us assume for the moment that $f\in C([0, 1])$ is a continuous function. Then the integrand $g_{s',s}(x):=|f|_{s'}(s'x)- f|_s(sx)|^2$ is dominated by $4\|f\|_\infty^2$
and $g_{s',s}(x)\to 0$ pointwise when $s'\to s$ in view of the continuity. As a consequence the application of the Lebesgue's dominated convergence theorem assures that $b_f(s',s)\to 0$ when $s'\to s$. To sum up we have deduced that $\|(D_{s'}-D_s)f\|\to 0$
 when $s'\to s$ for every $f\in C([0, 1])$. Since continuous functions are dense in $L^2$ functions one can conclude that the map $s\mapsto D_s$ is indeed strongly continuous. The last part of the claim follows from $\Phi(0,P)=\phi_1(P)=P$ and 
 Lemma \ref{lemm_lim_0} which provides $\Phi(1,P)=\lim_{t\to 0}\phi_t(P)=0$.
\qed

\begin{remark}[Deformation retraction]\label{rk_def}
In the jargon of topology the map $\Phi$ introduced in Proposition \ref{prop_def_ref} is a \emph{deformation retraction} of the space 
${^{\rm w}\bb{G}_{\bullet}}$ to the singleton $\{0\}$ \cite[Chapter 0]{hatcher-02}. Since  the map $\Phi$ doesn't   increase the trace of projections (as for Corollary \ref{cor_insid_N}), it follows also that  $\Phi$ retracts each subspace ${^{\rm w}\bb{G}_{\bullet,\le N}}$ to $\{0\}$. Therefore Proposition \ref{prop_def_ref} implies that the spaces ${^{\rm w}\bb{G}_{\bullet}}$ and ${^{\rm w}\bb{G}_{\bullet,\le N}}$ are homotopy equivalent to the singleton $\{0\}$. In other words, they are contractible.
\hfill $\blacktriangleleft$
\end{remark}


\begin{thebibliography} {[RMCPV]}
\frenchspacing \baselineskip=12 pt plus 1pt minus 1pt


\bibitem[AM]{abbondandolo-majer-09} 
{Abbondandolo, A.; Majer, P.}:
{\sl Infinite dimensional Grassmannians}. 
J. Op. Theory {\bf 61}, 19-62 (2009)


\bibitem[AS]{Atiyah-segal-04} 
{Atiyah, M. F.; Segal, G.}:  
{\em Twisted K-theory}. Ukr. Mat. Visn. {\bf 1}, no. 3, 287-330
(2004). English transl., Ukr. Math. Bull. {\bf 1}, no. 3, 291-334 (2004) 


\bibitem[Atiy]{Atiyah-67} 
Atiyah, M. F.:  
{\em $K$-Theory}. CRC Press, 1967


\bibitem[BR1]{bratteli-robinson-87} {Bratteli, O.; Robinson, D. W.}: {\em  Operator Algebras and Quantum Statistical Mechanics 1. $C^*$- and $W^*$-Algebras, Symmetry Groups, Decomposition of States}. 
 Springer-Verlag, Berlin-Heidelberg, 1987


\bibitem[De]{denittis-25} De~Nittis,~G.: {\sl Topological phases of non-interacting systems: A general approach based on states}.
Lett. Math Phys. {\bf 115}, (2025) 


\bibitem[DR]{denittis-rendel-25} De~Nittis,~G.; Rendel,~S.G.: {\sl Topological description of pure invariant states of the Weyl $C^*$-algebra}.
{Rev. Math. Phys.}~\textbf{38}(05), (2026)


\bibitem[Dup]{dupre-74} 
{Dupr\'e, J. M.:} 
{\sl Classifying Hilbert Bundles}. 
J. Funct. Anal. {\bf 15}, 244-278 (1974)


\bibitem[EU]{espinoza-uribe-14} 
{Espinoza, J.;   Uribe, B.:} 
{\sl Topological properties of the unitary group}. 
JP J. Geom. Topol.  {\bf 16}, 45-55 (2014)


\bibitem[Hat1]{hatcher-02} Hatcher,~A.: {\em  Algebraic Topology}. Cambridge University Press, Cambridge,  2002


\bibitem[Hat2]{hatcher-17}
{Hatcher, A.}:
{\em Vector Bundles and $K$-Theory}\\ 
E-print: {\tt https://pi.math.cornell.edu/~hatcher/VBKT/VB.pdf}, 2017


\bibitem[Huse]{husemoller-94} Husemoller,~D.: {\em  Fibre bundles}. Springer-Verlag, New York, 1994


\bibitem[Kuip]{kuiper-65}
{Kuiper N.}:
{\sl The homotopy type of the unitary group of Hilbert space}. 
Topology {\bf 3}, 19-30 (1965). 


\bibitem[Miln]{Milnor-56} 
{Milnor, J.}: 
{\sl Construction of universal bundles II}. 
{Ann. of Math.}~\textbf{63}, 430-443 (1956)


\bibitem[MMS]{matumoto-minami-sugawara-84} 
{Matumoto, T.; Minami, N.; Sugawara, M.}: 
{\sl On the set of free homotopy classes and Brown’s construction}. 
{Hiroshima Math. J.}~\textbf{14}, 359-369 (1984)


\bibitem[MS]{milnor-stasheff-74}
{Milnor, J. W.; Stasheff, J. D.}
{\em Characteristic Classes}
Princeton University Press, 1974


\bibitem[Pala1]{palais-65} 
{Palais, R. S.}: 
{\sl On the homotopy type of certain groups of operators}. 
{Topology}~{\bf 3},  271-279 (1965)


\bibitem[Pala2]{palais-66} 
{Palais, R. S.}: 
{\sl Homotopy theory of infinite-dimensional manifolds}. 
{Topology}~{\bf 5},  1-16 (1966)


\bibitem[RS]{rudolph-schmidt-17}
{Rudolph, G.; Schmidt, M.}:
{\em Differential Geometry and Mathematical Physics Part II. Fibre Bundles, Topology and Gauge Fields}.
Theoretical and Mathematical Physics, Springer, 2017


\bibitem[Scho]{schottenloher-18} 
{Schottenloher, M.}:
{\sl The Unitary Group in Its Strong Topology}. Adv. Pure Math. {\bf 8}, 508-515 (2018)


\bibitem[Shu]{shubin-96} 
{Shubin, M. A.}:
{\sl Remarks on the Topology of the Hilbert Grassmannian}. Amer.Math. Soc. Transl. {\bf 175}, 191-198 (1996)


\bibitem[Simo]{simon-05} 
{Simon,~B.}: 
{\em Trace Ideals and Their Applications}. 
AMS, 2005


\end{thebibliography}
\end{document}